\documentclass[rmp,twocolumn,showkeys,superscriptaddress]{revtex4}
\usepackage{hyperref}
\usepackage{graphicx}
\usepackage{times}

\newcommand{\figurewidth}{8.4cm}
\newcommand{\widefigurewidth}{17.5cm}
\newcommand{\ESM}{\emph{ESM}}

\begin{document}
\title{Epistasis can lead to fragmented neutral spaces and  contingency in evolution}
\author{Steffen Schaper}
\email{steffen.schaper@physics.ox.ac.uk}
\affiliation{Rudolf Peierls Centre for Theoretical Physics, University of Oxford, UK}

\author{Iain G. Johnston}
\affiliation{Department of Physics and CABDyN Complexity Centre, University of Oxford, UK}
\affiliation{Oxford Centre for Integrative Systems Biology, Department of Biochemistry, University of Oxford, UK} 

\author{Ard A. Louis}
\email{ard.louis@physics.ox.ac.uk}
\affiliation{Rudolf Peierls Centre for Theoretical Physics, University of Oxford, UK}

\begin{abstract}

\noindent In evolution, the effects of a single deleterious mutation can sometimes be compensated for by a
second mutation which recovers the original phenotype. Such epistatic interactions have implications for the structure
of genome space -- namely, that networks of genomes encoding the same phenotype may not be connected by single
mutational moves. We use the folding of RNA sequences into secondary structures as a model genotype-phenotype map and
explore the neutral spaces corresponding to networks of genotypes with the same phenotype. In most of these networks, we
f\/ind that it is not possible to connect all genotypes to one another by single point mutations. Instead, a network for
a phenotypic structure with $n$ bonds typically fragments into at least $2^n$ neutral components, often of similar size.
While components of the same network generate the same phenotype, they show important variations in their properties, most
strikingly in their evolvability and mutational robustness. This heterogeneity implies contingency in the evolutionary process. 

\end{abstract}

\maketitle

\section{Introduction}

\par The course of evolution is shaped by  the complex interaction between random mutations that change
genotypes and  natural selection that acts on variation between phenotypes.  Progress in evolutionary theory is thus predicated on
gaining further understanding of the structure of genotype-phenotype (GP) maps \cite{alberch1991gpmap}. These mappings
exhibit many non-trivial properties. For example, as emphasised by Kimura \cite{kimura1985book}, many mutations are
neutral -- they do not appreciably change the phenotype or f\/itness -- leading to a many-to-one redundancy in the
transformation from genotypes to phenotypes that has profound consequences for evolution. In the {\em neutral theory of
evolution}, genetic changes that are invisible to selection \cite{ohta1973} can build up over time and may  constitute the majority
of mutations in an evolutionary lineage. Evidence for the abundance of neutral mutations can be found, for example, in
homologous proteins that differ in sequence, but perform the same or very similar tasks in different organisms
\cite{kimura1985book}.

\par {\em Epistasis} describes another important property of GP maps:  the phenotypic effect of a genetic
change at a single locus may depend on the values of other genetic loci. That such dependencies should
exist is not at all surprising. Given the many multi-scale physical processes involved in translating a genotype into a
phenotype, it is rather the absence of epistasis that might be expected to be the exception to the rule.

\par Recent advances in high throughput techniques and in bioinformatics have facilitated many new experimental studies
of epistasis.  For example,  Lunzer et al. \cite{lunzer2010epistasis} studied  the  leuB gene that codes for  
$\beta$-isopropylmalate dehydrogenase in both {\em E. coli} and {\em P. aeruginosa}.  These two homologous proteins
differ at 168 positions, but when the mutations were implemented  individually in {\em E. coli}, 63 of them were found
to be individually deleterious, suggesting rampant epistasis, since their overall effect is neutral. Other recent
studies have found large-scale epistasis in HIV-1 virus genes \cite{dasilva2010HIV,hinkley2011HIV}, and in mitochondrial
transfer RNA from eukaryotes \cite{kondrashov2010RNA}. These three examples constitute only a very small snapshot of a
much larger body of literature that suggests that epistasis is widespread throughout the living world
\cite{phillips2008epistasis,romero2009review}.
  
\par The ubiquity of epistasis also implies  that neutral evolution can play a key role in facilitating the genotypic
background that allows evolution to climb an adaptive peak \cite{wagner2008neutralism}:  A set of mutations can be
initially neutral, but when the environment  or the genotype changes, they may either be adaptive themselves, or bring
a population closer to potential adaptive innovations.  In other words, neutral  evolution may enhance evolvability, the
ability of an organism to facilitate heritable phenotypic changes \cite{pigliucci2008evolvability}. For example, in a
recent paper Hayden et al. \cite{hayden2011RNA} showed that allowing a population of ribozymes to accumulate neutral
mutations greatly increased the population's ability to  adapt to a new environment, and that this enhanced
evolvability could be traced to `cryptic' variation that arose neutrally.
  
 \par In the context of evolution it is helpful to quantify  epistasis in terms of the f\/itness that
selection can act on\footnote{In this paper we will ignore the potentially very complex mapping from phenotypes to
f\/itness, and simply assume that f\/itness can be ascribed to and differs between phenotypes.}
\cite{phillips2008epistasis}. Epistasis manifests in many different ways. In this paper we concentrate on just two of these.
Consider, for example, a simple two allele two locus system with alleles $a$ or $A$ at locus one, and $b$ or $B$ at
locus two.  If the transition from $ab$ to $AB$ increases f\/itness then  {\em sign epistasis}
\cite{weinreich2005epistasis} describes the situation where  {\em either} $aB$ or $Ab$ has a lower f\/itness than $ab$,
whereas {\em reciprocal sign epistasis} \cite{poelwijk2007landscape} occurs if {\em  both} intermediate genetic states
have a lower f\/itness than $ab$.   Sign epistasis constrains the potential pathways that evolution can take towards high
f\/itness phenotypes~\cite{weinreich2006paths},  whereas reciprocal sign epistasis is a necessary, but not suff\/icient,
condition for peaks in f\/itness landscapes \cite{poelwijk2011epistasis}. Even in this simple biallelic two locus system
one can imagine other epistatic scenarios \cite{weinreich2005epistasis,poelwijk2007landscape}, and the potential for
complexity increases greatly as more genetic loci are considered.

\par The considerations above frame the main question to be addressed in this paper: If epistasis constrains the
pathways of adaptive mutations, \textit{can it also constrain the potential for neutral mutations to facilitate
adaptation?}
 
\par Although epistasis can have many different consequences for neutral evolution, in this paper we will in particular
focus on the role of  {\em neutral reciprocal sign epistasis}: Consider our biallelic system -- if both $ab$ and $AB$
have the same f\/itness, but $Ab$ and $aB$ are unviable, then the only way to get directly from $ab$ to $AB$ is through
double mutations.  In this context, it is helpful to def\/ine  \emph{neutral networks} (NNs)
\cite{schuster1994neutralnetworks}: sets of genotypes that  share the same phenotype.  If we are in the regime of
strong selection and weak mutation, the main case we consider in this paper, then  double mutations will be very rare.
One consequence of this reciprocal sign epistasis will be that  an NN that contains $ab$ and $AB$ may be fragmented into
separate \emph{neutral components} (NCs). If the NN is fragmented into several NCs, this raises further questions like:
\textit{ Are these NCs homogeneous or heterogeneous? Does the potential for innovation depend on which NC a population
f\/inds itself in?}
  
\par There are many potential causes of neutral reciprocal sign epistasis.  For example, any mechanism that resembles a
lock and a key may need two compensatory mutations, one for the lock, and the other for the key, in order to restore
function.   In his classic paper on compensatory mutations,  Kimura \cite{kimura1985compensatory} considered the case
of two interacting amino acid sites for which a mutation in either amino acid is deleterious, but where a double
mutation can restore the function. Although these two sites are physically close in the folded state, they may be far
away along the protein backbone, and it is hard to be sure that a correlated set of mutations at other positions may
not allow the two sites to change by single mutational states. Thus, just as is the case for f\/itness peaks
\cite{poelwijk2011epistasis}, neutral reciprocal sign epistasis is a necessary, but not suff\/icient condition, for
disconnected NNs.

\par In this context, it is also important to remember that the GP map is typically characterised by very 
high dimensions, a property whose consequences have been of recent theoretical interest 
\cite{pigliucci2008fitness,gavrilets1997percolation,gavrilets2004book}. Briefly put, isolated f\/itness peaks are less
likely to occur in high dimensional landscapes; instead, long neutral ridges feature much more prominently. NNs can be
identif\/ied with these ridges, and by traversing these networks, populations can explore large proportions of genotype
space without having to cross f\/itness valleys.   Similar arguments  suggest that even when neutral reciprocal sign
epistasis breaks a pathway between two genetic conf\/igurations, there may nevertheless be other pathways that connect up
the NN.  Thus an investigation of NCs necessitates either a fairly complete description of the GP map, or
alternatively, a good enough understanding of local topology to ensure that an NN is disconnected.
  
\par For these reasons we concentrate in this paper on a computationally tractable and biologically motivated GP
mapping. RNA strands can fold into well-def\/ined three-dimensional structures driven by the specif\/ic bonding between
AU, GU and GC base pairs, as well as stacking interactions between adjacent bases.  The RNA secondary structure
describes the bonding pattern of a folded RNA strand of length $L$.  There exist eff\/icient and reliable algorithms that
predict secondary structure from primary sequence by minimising the free-energy. For the work presented here, we use
the \texttt{RNAfold} program, version $1.8.4$ \cite{ViennaPackage}.  This system describes a map from a genotype of
length $L$ to a phenotype that is characterised by the secondary structure. It  has been extensively studied,
generating many important insights into evolutionary theory \cite{schuster1994neutralnetworks,
reidys1997RNAneutralnetworks,fontana1998continuity,fontana2002evodevo,wagner2005book}. The RNA map has the advantage
that for modest values of $L$ one can perform an exhaustive enumeration, and from this completely characterise the
connectivity of the NNs \cite{gruener1996RNAexhaustive}.
 
\par The paper is organised as follows: In section \ref{sec:fragmentation} we establish  that the NNs of most RNA
secondary structure phenotypes are fragmented into disconnected NCs.  We identify an important source of this
fragmentation to be a particular kind of neutral reciprocal sign epistasis that arises from the biophysics of the GP
map: Converting a pyrimidine-purine base-pair (e.g. GC) into a purine-pyrimidine pair (e.g. CG) in an RNA stem
motif cannot proceed by single mutations without passing through an intermediate of a different structure. By exhaustive
enumeration of length $L=15$ RNA sequences, we can study detailed properties of the NCs. We establish that many NNs can
be split into multiple components with no particular NC being dominant. 
% Although we do not directly study evolutionary dynamics, we do 
We also show that the fragmentation of these NNs will be sustained under crossover moves, implying that our
results may be relevant for populations in both asexual and sexual regimes.
 
\par We next examine some consequences of this fragmentation of NNs in section \ref{sec:contingency}. We show that
the size of a given NC component correlates with a measure of its robustness to genetic mutations. Since a typical NN is
fragmented into multiple NCs of different size, this implies that the robustness of a given population will depend on
which NC it is on, and not only on its phenotype. Similarly, we f\/ind that the  number of phenotypes accessible
within one point mutation of the NCs, a measure of their evolvability \cite{wagner2008paradox},  varies signif\/icantly
between different NCs in a given NN. This heterogeneity leads us to conclude that the evolutionary fate of a population
is contingent on the NC it occupies in genotype space.  Finally, in section \ref{sec:discussion} we discuss our main
results, and look beyond the RNA secondary structure GP map to consider which conclusions may hold for
a wider class of systems in including gene regulatory networks, proteins and the genetic code.

\section{RNA neutral networks are fragmented}
\label{sec:fragmentation} 

\begin{figure}[!tb]
\centering
\includegraphics[width=\figurewidth]{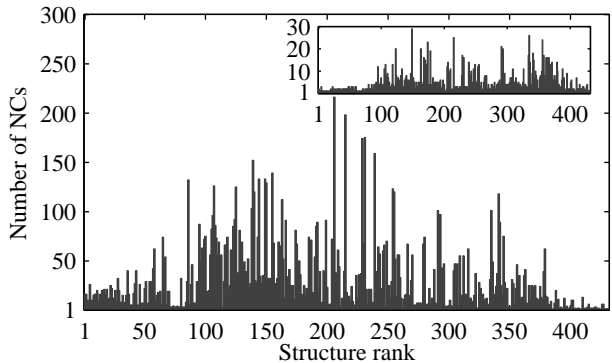}
\caption{\textbf{RNA neutral networks are split into many components.} The number of NCs that make up the NN of a given
structure is plotted against the size ranking of the structure, starting at $1$ for the largest NN. The inset of the
figure shows the consequence of allowing base pair exchanges as a fundamental evolutionary step. Data is for $L=15$.}
\label{fig:number_of_NCs}
\end{figure}

\par The structure of NNs in RNA have been extensively studied previously
\cite{schuster1994neutralnetworks,reidys1997RNAneutralnetworks,gruener1996RNAexhaustive, cowperthwaite2008ascent}. Here
we briefly repeat some key results of this earlier work that are relevant for our investigations (see also
Electronic Supplementary Material (\ESM), Table \ref{tab:OldResults}):

\begin{itemize}
  \setlength{\itemsep}{1pt}
  \setlength{\parskip}{0pt}
  \setlength{\parsep}{0pt}
  \item The number of structures is much smaller than the number of sequences $4^L$. The number of structures increases
  with $L$, but at a much slower rate than the number of sequences.
  \item The distribution of NN sizes is heavily skewed, that is a minority of the phenotypes occupy a majority of the
  genotypes.  For a given $L$,  NNs with more than average size are called \emph{large}, and the corresponding
  secondary  structures are said to be \emph{frequent}. While the absolute number of frequent structures increases with
  $L$, the fraction of sequences folding into frequent structures goes up, while the fraction of NN's that are large
  goes down.
  \item The fraction of sequences that fold into the trivial structure (that is the structure that has no bonds)
  decreases with $L$.
\end{itemize}

\par The connectivity of NNs can be studied under the simplifying assumption that a network is made up of randomly
chosen points on a genotypic hypercube.  Analyses using graph theory then suggest that larger NNs are are likely to be
fully connected, while small ones are likely to be fragmented \cite{reidys1997percolation,gavrilets2004book}.

\par For RNA secondary structure, however, it is important to also take the biophysics of bonding into account.  In
principle, each bond can be formed by one of six different nucleotide pairs: GC, CG, AU, UA, GU and UG. Point
mutations can  potentially connect GC$\leftrightarrow$ GU $\leftrightarrow$ AU and CG $\leftrightarrow$ UG
$\leftrightarrow$ UA, but these two subspaces cannot be connected together by point mutations without breaking a bond. 
This type of neutral reciprocal sign epistasis suggests that  for a structure with $n$ bonds we can expect on the order
of $2^n$ disjoint sets of compatible sequences\footnote{There may be other causes of fragmentation, and bonds vary in
energy, so not all possible combinations will lead to the same secondary structure.}. This argument is independent of
sequence length, and given that longer sequences may generate structures with more bonds, we expect the average number
of NCs per NN to grow with $L$ (see \ESM, Table \ref{tab:Components}).

\par We therefore predict that virtually all NNs for RNA secondary structure should be fragmented. By contrast, if
double mutations (base-pair swaps) are allowed, then the results from random graph theory give a good estimate of the
connectivity of an NN \cite{reidys1997percolation}.   But, in nature, base-pair swaps are expected to be very
rare \cite{kondrashov2010RNA}. While the fact that RNA secondary structure NNs are not fully connected has been widely
acknowledged in the literature \cite{gruener1996RNAexhaustive,cowperthwaite2008ascent,vanNimwegen1999robustness},
the potential consequences of this fragmentation have not yet been fully explored.

\par In order to determine the connectivity of the NNs, we start from a random sequence
in the NN and follow all neutral mutations that can be accessed \cite{gruener1996RNAexhaustive}. Sequence space grows
exponentially with length $L$, so this exhaustive approach is only feasible for relatively short sequences; we will
mainly present  results for sequence length up to $L=15$, but will also consider other lengths where appropriate.  As
has been done in many other studies \cite{fontana1998continuity,cowperthwaite2008ascent}, we ignore the trivial
structure with no bonds\footnote{In systems where folded structures have an adaptive advantage, it is likely that the
completely unfolded strand has very low f\/itness, and so can be ignored.  There is also a practical reason for this
choice. The trivial structure is much more frequent for small $L$ than for large $L$, and so it could affect the
applicability of our results for much longer structures.}.

\par Our $L=15$ system has $431$ distinct secondary structures (at a folding temperature of $37^\circ$C) of which $86$
or about $20\%$ are large. The large structures cover $93\%$ of the folding sequences. By exhaustively searching 
through all of sequence space, we are able to identify all  $12526$ components, so that there are on average about $29$
components per neutral network. Figure \ref{fig:number_of_NCs} shows how
these are distributed among the different NNs.  The largest number of components is $216$ for a relatively infrequent structure ranked
$206$th, and only a few small structures  have a single NC (the largest has rank 333). We summarise the data in
\ESM, Tables \ref{tab:OldResults},\ref{tab:Components} and Figs.
\ref{fig:nn_skewness},\ref{fig:nc_skewness}. The NCs can be individually ordered, and are even more skewed than the
NNs. Overall, $1120$ NCs (less than $10\%$) are larger than average, but together they cover $95\%$ of non-trivial
genotype space.

\par By analogy to NNs we call an NC large if its size is more than the average in its NN. Most NNs contain several
large NCs (see \ESM, Fig. \ref{fig:large_ncs}). Rather than being dominated by one NC, we observe that for most
phenotypes there are many large NCs. The number of large NCs in an NN is strongly correlated with $2^n$ where $n$ is
the number of bonds in the corresponding secondary structure ($r=0.74$). In contrast, there is hardly any correlation
between the number of large NCs and NN size ($r=-0.01$).

\begin{figure}[!tb]
\centering
\includegraphics[width=\figurewidth]{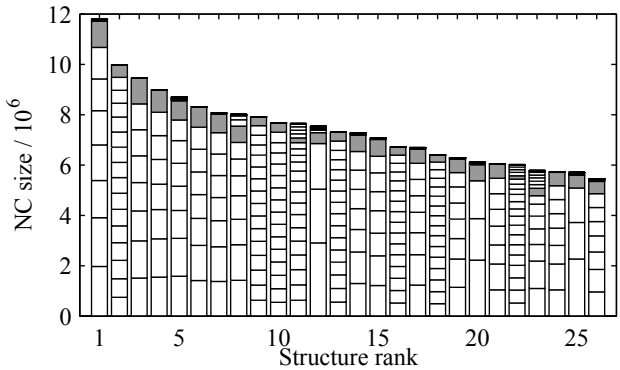}
\caption{\textbf{RNA NNs contain several large NCs.} The $26$ largest NNs shown here cover just over $50\%$ of folding
genotype space. For each NN, the sizes of all its NCs are shown. The shaded NC denotes the $2^n$th NC ($n$ is the number
of bonds), and the thick black lines at the top of a stack indicate the existence of signif\/icantly smaller NCs. The 12
most abundant secondary structures are shown in \ESM, Fig. \ref{fig:largest_structures} and Table
\ref{tab:largest_structures}.}
\label{fig:nc_size_scatter}
\end{figure}
 
\par In Figure \ref{fig:nc_size_scatter} we show the size of the components for the largest $26$ NNs that, together,
cover over $50 \%$ of the folding genotypes. Most networks have more than the $2^n$ components we expect due to the
biophysical argument given above; nonetheless the $2^n$ largest NCs are generally very similar in size, and much larger
than all the smaller NCs of the network. Note that the largest NC is for an NN that is ranked $12$th by overall size, 
and more generally that the size of these largest $26$ NNs is not a reliable guide to the average size of the large
NCs. 

\par If in addition to the point mutations, we also allow base-pair swaps, then the number of components drops
signif\/icantly. In particular, as predicted by random graph theory \cite{reidys1997percolation}, the majority of large
NNs are  dominated by a single giant NC. This big difference, caused by introducing base pair swaps, strongly
suggests that the NN fragmentation we observe under point mutations arises from the simple neutral
reciprocal sign epistasis mechanism we identif\/ied above.

\par While single point mutations cannot connect up the NCs, one may consider whether crossover moves may do so.  In
that context it is helpful to consider Kimura's analogy  to a lock-key system \cite{kimura1985compensatory}: A change
in the lock makes it necessary for the key to be changed accordingly. Crossing over one lock-key setup with another can
only be successful in two cases.  First, if the lock and key originate from different parents, successful offspring
will arise only if the parents are compatible. Second the lock and key may originate from the same parent; this
requires that crossover arise at special points in the sequence to ensure a matching lock and key.

\par In RNA, the f\/irst case means that both parental sequences belong to the same NC and that the offspring consequently
stay in that NC. The second case is possible only if the point of crossover is outside the looped region of the stem
that is incompatible in the parent sequences. This is illustrated in \ESM, Fig.
\ref{fig:crossover}. Under the second condition, crossover can put offspring onto NCs that are distinct from either
parent; however, such crossover only allows to explore a small subset of all possible NCs in an NN.  Even this limited
exploration is predicated on the population being distributed on multiple NCs in the f\/irst place, but this cannot be
achieved without compensatory mutations. It is worth noting that crossover slows down the rate of f\/ixation of
compensatory mutations \cite{kimura1985compensatory}.

\par So far, we have shown that under fairly general conditions, the NNs of RNA secondary structure are
fragmented into many NCs. This fact raises the following question: \textit{Are the different NCs similar or
heterogenous in their properties?} 

\section{Neutral components shape evolutionary trajectories}

\label{sec:contingency}

\subsection{Robustness increases with component size}

\par The robustness to genetic change has been widely studied in the context of NNs. In particular, van
Nimwegen et al. have shown how the robustness of an evolving population depends on the structure of the underlying
neutral space \cite{vanNimwegen1999robustness}. While the dynamic properties of a population depend also on its size and
mutation rate, we consider here only the effect of the structure of the NC. To this end, we
def\/ine the mutational robustness of a genotype as the fraction of mutations that leave the phenotype unchanged.
% $R = \frac{\mu_N}{3 L}$, where $\mu_N$ is the number of possible mutations that do not change the phenotype, and $3 L$
% is the total number of possible mutations that can be made to an RNA strand of length $L$. 
In analogy to \cite{wagner2008paradox} we calculate
the robustness of an NC by averaging the genotypic robustness 
% $R$ 
of all genotypes in the NC. This
measure gives the expected average robustness of a monomorphic population evolving on the NC 
\cite{vanNimwegen1999robustness}.

\par
\begin{figure}[!tb]
\centering
\includegraphics[width=\figurewidth]{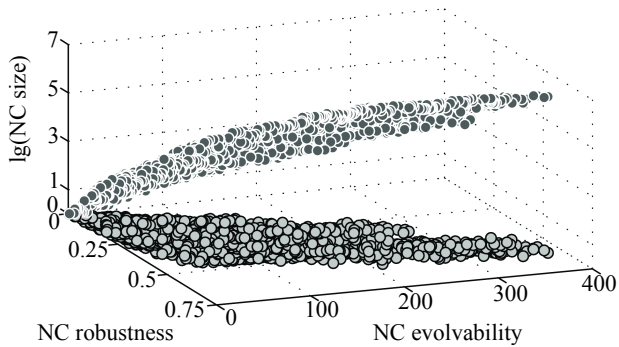}
\caption{\textbf{Robustness and evolvability increase with NC size.} The evolvability counts the number of
phenotypes that can be reached from an NC; robustness is the average probability that a mutation results in a genotype
in the same NC. The projection into the robustness-evolvability plane illustrates their positive
correlation.}
\label{fig:robustness_evolvability}
\end{figure}

 In agreement with earlier results based on sampling techniques \cite{wagner2008paradox} we observe a clear
positive correlation between mutational robustness of an NC and its size ($r=0.47$), as illustrated in Fig.
\ref{fig:robustness_evolvability}.  
Hence, the larger the NC,  the more likely individuals are to pass their phenotype
on to their offspring after a random mutation.
Given the large heterogeneity of NC sizes comprising a given NN,  these
results suggest that robustness estimates based on the NN as a whole will not be representative of the robustness
experienced by a population conf\/ined on a given NC. For example, if a population is restricted to a small component of
a very large NN, the effective mutational robustness will be (much) lower than that estimated for the NN as a whole.

\subsection{Evolvability varies between components of the same phenotype}

\begin{figure*}[!htb]
\centering
\includegraphics[width=\widefigurewidth]{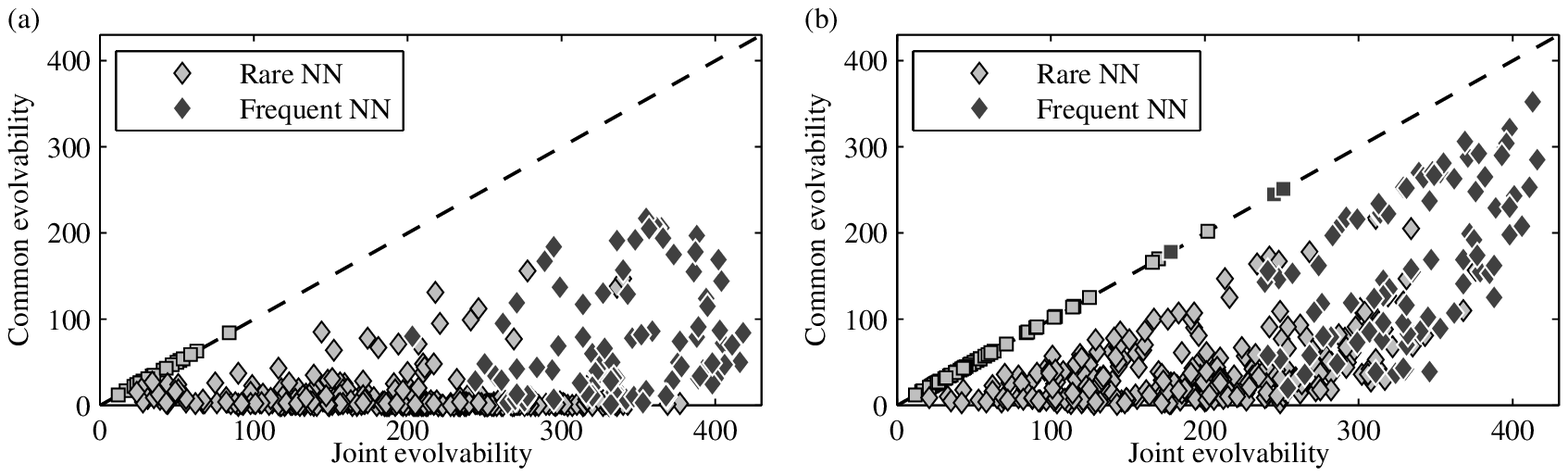}
\caption{\textbf{Phenotypic neighbourhoods are heterogeneous among NCs in the same NN.} \textbf{(a)} Here the joint and
common evolvability  are shown considering \emph{all} NCs of each NN. Square markers
indicate fully connected NNs, for which joint and common evolvability trivially coincide. \textbf{(b)} Here only the
large NCs in each NN were used for the calculations. Now square markers indicate NNs with only one large NC, for
which again the equality is trivial. In both panels the black dashed line indicates the equality of joint and common
evolvability.}
\label{fig:joint_common_evolvability}
\end{figure*}
	
\par The evolvability of a population is related to its ability to produce heritable phenotypic change
\cite{pigliucci2008evolvability}. One might naively think that the more robust a phenotype is to mutations, the
harder it is for mutations to generate novelty. However, this argument ignores the ability of neutral
exploration to pave the way for future adaptive innovations \cite{wagner2008neutralism}. Wagner has
proposed a  proxy measure of evolvability that counts the number of phenotypes $E_\mathcal{P}$ that can be reached by a
single mutation from a given NN \cite{wagner2008paradox}.  He showed that this measure also correlates positively with
the size of an NN, and argued that phenotypes with larger NN may be simultaneously more robust and more evolvable. But
if the NNs are fragmented into separate NCs, then it is in fact the NC robustness and evolvability that matter to a
population, and not the properties of the whole NN, which it cannot access by point mutations. Nevertheless, we f\/ind
that the  average robustness and evolvability of individual NCs are positively correlated (Fig.
\ref{fig:robustness_evolvability}, \ESM, Figs. \ref{fig:robustness} and \ref{fig:evolvability}) just as was found for
NNs. We can thus qualify's Wagner's result \cite{wagner2008paradox}: Robustness and evolvability are not so much
correlated at the level of phenotypes (NNs), but rather the correlation holds \textit{at the level of an NC}
(our results yield $r=0.81$), which can vary strongly within one NN.

\par Thus different populations with the same phenotype may exhibit signif\/icantly different  evolvabilities.  These
differences can be further quantif\/ied with the following def\/initions:
\begin{eqnarray}
	E_\mathcal{P}^{(j)} & = & \left| \bigcup_{c \in \mathcal{P}} E_c \right| 
	\label{eq:jointevo} \\
	E_\mathcal{P}^{(c)} & = & \left| \bigcap_{c \in \mathcal{P}} E_c \right|
	\label{eq:commonevo} 
\end{eqnarray}
where $E_c$ is the set of phenotypes that can be reached by a single mutation from NC $c$. Thus the \emph{joint
evolvability} (\ref{eq:jointevo}) counts the number of structures that can be reached from \emph{at least one} NC in the
NN, while the \emph{common evolvability} (\ref{eq:commonevo}) counts the structures available from \emph{all} NCs. The
comparison of these two properties reveals signif\/icant heterogeneity in the phenotypic neighbourhoods of NCs in the same
NN (see Fig. \ref{fig:joint_common_evolvability}a). In fact the joint and common evolvability are only identical for
those small NNs that are fully connected. For most NCs, a population will only be able to access a restricted subset
of the entire NN's neighbouring phenotypes: Averaged over all NNs, $F \equiv E^{(c)}/E^{(j)}=0.14$. There is no
signif\/icant correlation of $F$ and NN size: $r=0.04, p=0.41$.

\par We can further explore this heterogeneity by restricting our analysis to the large NCs only. We expect the
differences to diminish because the joint evolvability has a lower bound given by the most evolvable NC
while the common evolvability cannot be larger than for the least evolvable NC, which is typically very small. 
As Fig. \ref{fig:joint_common_evolvability}b shows, the ratio of common to joint evolvability decreases when only
large NCs taken into account: $F_{large}=0.37$ with a weak correlation with NN size: $r=0.17, p=4.5\times
10^{-4}$. We note that some phenotypes are only accessible from small NCs so the joint evolvability decreases slightly
(on average by about $10\%$). In \ESM, Fig. \ref{fig:joint_common_large} we restrict the phenotypes further to just
those that are large -- the same general results hold. Finally, we can ask what fraction of the joint evolvability are
accessible on average from a single NC. If we consider only the large NCs of the frequent phenotypes, this fraction is
on average $76\%$ (in agreement with \cite{cowperthwaite2008ascent}), while averaging over all NCs in all NNs brings
this down to $42\%$ (see also \ESM, Fig. \ref{fig:rel_evo}). Instead of requiring large NCs to be
greater than the average NC in their NN, we also employed an entropy-based criterion and obtained qualitatively similar
results (\ESM, Sec. \ref{sec:entropy} and Fig. \ref{fig:evo_by_entropy}).

\par It is important to consider whether this discrepancy is an artefact of the relative short sequences we
study. Answering this question by exhaustive enumeration is unfeasible. Instead, we employed a sampling
technique (\ESM, Sec. \ref{sec:l20}) for sequences of 20 nucleotides. We f\/ind that the heterogeneity
between NCs becomes even more pronounced as the sequence length increases (\ESM, Fig. \ref{fig:sampling_evo}).

\par Taken together, we have arrived at a key result: the potential for future innovation does not only
depend on the current phenotype, but also on which NC a population occupies. The fact that different NCs provide access
to different new phenotypes suggests a new mechanism for  contingency in evolution. A dynamic setting in which this may
be particularly important is a polymorphic population with genotypes from two (or more) NCs. If environmental changes
are suff\/iciently rapid (that is faster than genetic drift), this could drive parts of the population to different
phenotypes, potentially aiding diversif\/ication at the phenotype level \cite{gavrilets2004book}.

\section{Discussion}
\label{sec:discussion}

\par  We have shown how neutral reciprocal sign epistasis in RNA leads to fragmentation of NNs into multiple
components.   For many of the NNs, no one component dominates.  Moreover, the components are heterogeneous, so that
different populations with the same phenotype, but different NCs, may show large variations in robustness and
evolvability.

\par These inferences were possible because of the tractability of the GP map between an RNA sequence and
its secondary structure.  An obvious question is whether our results extend to other maps.   Boldhaus and Klemm
\cite{boldhaus2010GRNcomponents} studied a coarse-grained Boolean threshold dynamics model \cite{li2004yeast}  for the
regulatory network of the yeast cell cycle and identif\/ied nearly half a billion functional NCs, ranging in size between
$6.1\times 10^{24}$ and $4.4\times 10^{26}$ genotypes. Interestingly, the wild type network is part of one of the
smaller NCs. It contains networks which are quite sparse and noise-resilient, indicating that there are secondary
aspects in the performance of the network which can be selected for.  This example also shows heterogeneity in
the properties of NCs. One caveat is that the point mutations were in an abstract space with discretised interactions. 
It is not yet clear how a more realistic model of mutations would affect the NCs.

\par Recent experimental reconstructions of f\/itness landscapes may also open up avenues to study NCs.  For example, in
an important paper, Weinreich et al. \cite{weinreich2006paths}  characterized all $32$ combinations of $5$ mutations
that together increase resistance to a particular antibiotic by a factor of about $10^5$. By measuring the resistance of
each possible combination, they produced a phenotype landscape; in examining their data, we
found that this landscape also contains several NCs (see \ESM, Fig. \ref{fig:weinreich}). 

\par There are two important caveats to this f\/inding: First, the resistance scale used in the experiment is relatively
coarse. Thus the neutrality in this landscape may even be broken by relatively small populations. In general, we stress
that neutrality is always an effective statement, depending on population size \cite{ohta1973}.Second, we cannot
exclude the existence of neutral connections that were outside the scope of the experiment. Excluding such paths by
exhaustively cataloguing all possible mutations would be prohibitive. Progress can be made by studying the biophysics
of a GP map, and looking for examples of lock and key type systems. In proteins, binding sites may be potential
candidates~\cite{kimura1985compensatory}.  However, as reviewed by Poelwijk et al. \cite{poelwijk2007landscape}, even
lock and key systems can sometimes evolve in subtle ways through single mutations.

\par Another system to consider is the genetic code. It is interesting to note that with the exception of serine, all
sets of codons coding for a particular amino acid can be reached by single synonymous point mutations. However, serine
has two NCs, one made up of AGU and AGC, and the other of UCU, UCC, UCA and UCG. Given that serine often plays a key
role in active sites in proteins, it may be that it cannot easily be neutrally replaced by another amino acid, so that
these two NCs may indeed be separate in nature. It is noteworthy that this high NN connectivity is extremely unlikely to
arise in a random genetic code with the same degeneracy as the universal code (\ESM, Fig. \ref{fig:genetic_code}). As
robustness correlates with NC (and not NN) size, this striking degree of NN connectivity may be a by-product of
selection for other properties such as robustness of the genetic code to point mutations or translation errors
\cite{wagner2005book}.

\par We have focussed on the approximation of strong selection and weak mutations where double mutations are
excluded. However, compensatory mutations can occur if the f\/itness penalties are weak, or if mutation rates are high.
Measuring f\/itness is notoriously diff\/icult, but a recent study of compensatory mutations in mitochondrial  transfer RNA
estimates that transitions from GC to AU  may occur through low f\/itness GU and AC intermediates
\cite{kondrashov2010RNA}. By contrast, switches like AU$\leftrightarrow$UA, GC$\leftrightarrow$CG and
AU$\leftrightarrow$CG, each of which requires a transversion, were found to be very rare. These results suggest that
in nature we should expect fragmented neutral spaces in RNA to be common.

\par Nevertheless, a suff\/iciently large population and/or high mutation rate can lead to
a regime in which NCs are effectively connected, so that evolutionary dynamics may be less sensitive to the effects of
NN fragmentation. In the opposite limit of small populations and/or mutation rates, the average spread of a population
in genotype space can be much smaller than the size of many NCs, implying that the local NC structure becomes more
important. For a f\/ixed mutation rate, there will thus be a crossover in the effect of NCs on evolutionary dynamics with
increasing population size. More generally, the dependence of of evolvability and neutral space exploration on dynamic
parameters is an important issue that we plan to address in a future publication.

\par Our analysis has only considered the local phenotypic neighbourhood of individual NCs. Over longer
evolutionary timescales, populations evolve from one phenotype (and hence NC) to another and traverse the phenotypic
landscape. In order to understand the importance of landscape structure on such long timescales, it is necessary to
study not only accessible phenotypes, but also the connectivity among NCs, which will be the focus of future studies.

\par In conclusion then, we have focussed on one striking effect of epistasis on neutral evolution, namely
the fragmentation of neutral spaces. The heterogeneity of the resultant NCs is important both conceptually and in
practice: Properties such as the robustness and evolvability of an evolving population may not only depend on its
phenotype, but also on which NC of that phenotype the population occupies. This sensitivity may lead to contingency in
evolution: The evolutionary trajectory of a population depends not only on the occurrence of random mutations, but also
on the possible innovations that are available to the NCs it happens upon.

\onecolumngrid
\newpage

\begin{center}
	{\huge Electronic Supplementary Material}
\end{center}

\setcounter{section}{0}
\setcounter{figure}{0}
\renewcommand{\thefigure}{S\arabic{figure}}
\renewcommand{\thetable}{S\arabic{table}}
\renewcommand{\thesection}{S\arabic{section}}

\section{Introduction}

\par The state space for evolving individuals is genotype space. For DNA (or RNA) sequences of length $L$, genotype
space contains $4^L$ discrete points, each one corresponding to a unique sequence. Each point can be linked to $3L$
\emph{one-mutational neighbours} which differ from the original genotype at only one nucleotide in the sequence. These
connections between genotypes create a generalized hypercube in $L$-dimensional space.

\par For the purpose of visualisation, we may think of the mapping from genotypes to phenotypes of as a colouring of
the hypercube. Then all the vertices (genotypes) with the same colour (phenotype) make up a neutral network (NN).
Neutral components (NCs) are sets of genotypes that are connected on the hypercube and share the same phenotype. 

\par
\begin{figure}[!ht]
\centering
\includegraphics[width=\figurewidth]{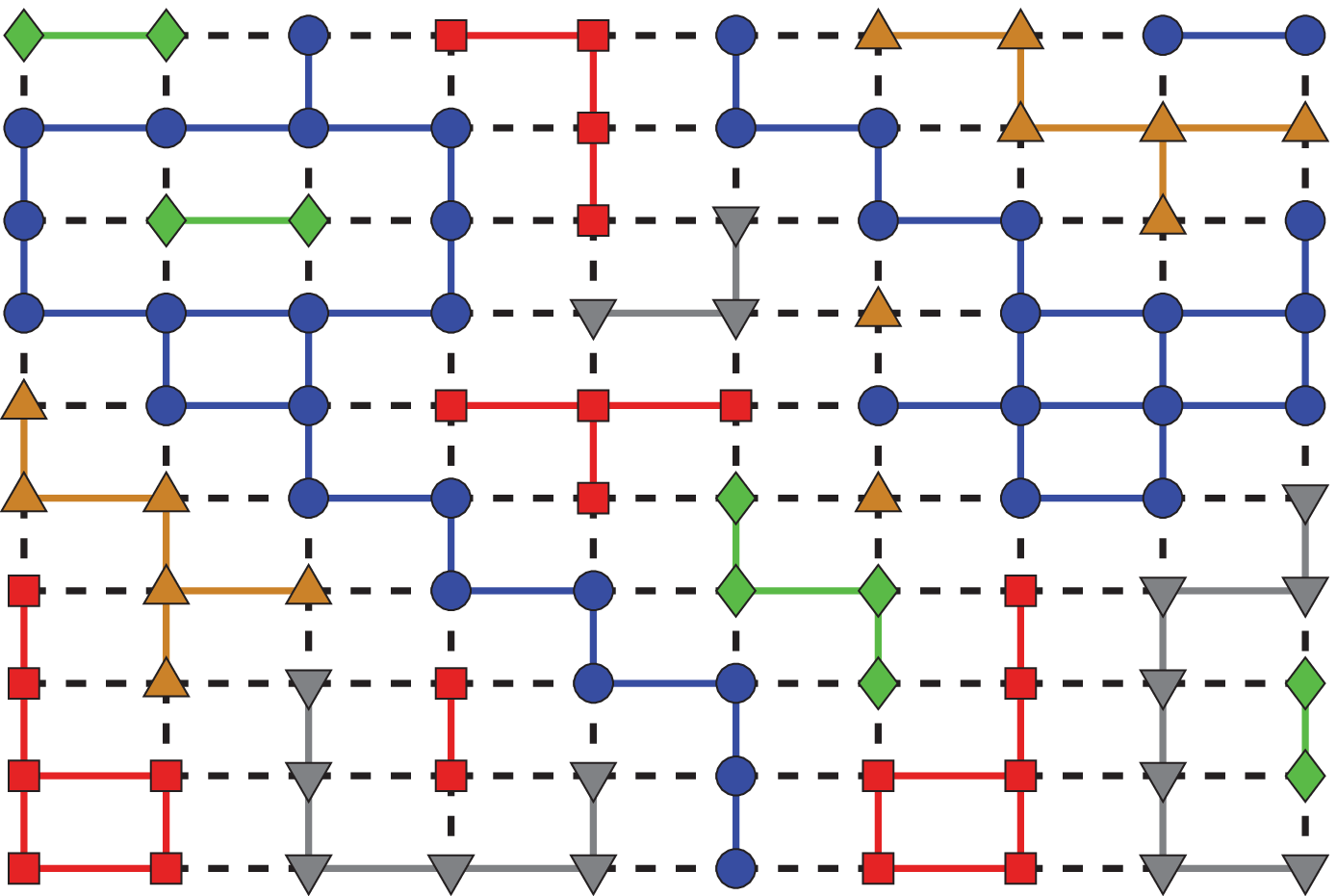}
\caption{\textbf{An illustration of genotype space.} The picture shows a simplif\/ied genotype space.  Each marker
corresponds to a genotype, phenotypes are coded for by shape and colour. Solid lines indicate neutral mutations, black
dashed connections are non-neutral mutations. This figure illustrates how neutral sets may be fragmented into separate
neutral networks. Note how different networks for the same phenotype differ in what other phenotypes can be reached by
single point mutations. In interpreting such pictures, it should be kept in mind that  real genotype spaces have much
higher dimensionality and no boundaries so that all nodes have the same number of neighbours. }
\label{fig:genotype_space}
\end{figure}

 It is hard to produce an accurate low-dimensional representation of the genotype hypercube. In Figure
\ref{fig:genotype_space}, we show a simplif\/ied picture. It is intended to illustrate the existence of neutral networks
and their components. In biologically relevant genotype spaces, the vertices have many more neighbours; in addition, there
are no boundaries on the hypercube.

\newpage\par\qquad
\section{RNA Neutral networks are fragmented}

\par Due to its computational tractability and biological relevance, the folding of RNA sequences into secondary
structures is a widely studied GP map \cite{schuster1994neutralnetworks,
gruener1996RNAexhaustive, reidys1997RNAneutralnetworks, fontana1998continuity, fontana2002evodevo,
wagner2008paradox, cowperthwaite2008ascent}. In Table \ref{tab:OldResults} we provide data on some well-known
characteristics of this map, and the scaling of these properties with sequence length $L$. In particular, the total
number of different secondary structures $n_S$ increases exponentially with $L$. Yet this increase is slower than the
expansion of sequence space as a whole (which grows as $4^L$), so that the average neutral network size $\langle V_S
\rangle$ also increases with $L$. We compute this average with the trivial structure excluded: this structure is
extremely frequent for the short sequences which we study, but it is clear that its abundance $r_{triv}$ decreases
quickly as $L$ increases. 

\par
\begin{figure}[!ht]
\centering
\includegraphics[width=\figurewidth]{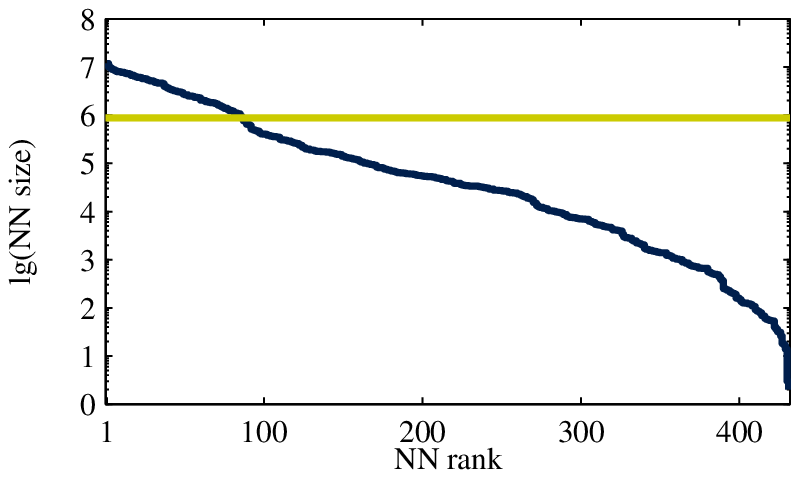}
\caption{\textbf{The distribution of NN sizes is skewed.} The figure shows the size of each NN against its rank,
starting at rank $1$ for the most frequent structure. The yellow line indicates the average NN size; only $86$ NNs
(that is, $20\%$) are larger than this. These large NNs contain $93\%$ of all folding sequences (cf. Table
\ref{tab:OldResults}).}
\label{fig:nn_skewness}
\end{figure}

 Figure \ref{fig:nn_skewness} shows the distribution of NN sizes, by which we just mean
the number of genotypes in the respective NN. The size distribution is strongly skewed: A few structures are frequent
while most structures are rare. To be precise, we call a secondary structure frequent if its NN is larger than the
average NN. Table \ref{tab:OldResults} shows that the absolute number of frequent structures $n_{freq}$ increases with
sequence length, while the fraction of structures that are frequent decreases for longer sequences. Nonetheless the
fraction of genotypes that map to one of the frequent phenotypes $r_{freq}$ grows with $L$.

\begin{table*}[!htb]
\centering
\begin{tabular}{|r|r|r|r|r|r|r|}
\hline
\multicolumn{1}{|c}{$L$} & \multicolumn{1}{|c}{$n_S$} & \multicolumn{1}{|c}{$r_{triv}$} & \multicolumn{1}{|c}{$\langle
V_S\rangle$} & \multicolumn{1}{|c}{$n_{freq}$} & \multicolumn{1}{|c}{$n_{freq}/n_S$} & \multicolumn{1}{|c|}{$r_{freq}$}
\\
\hline
$12$ & $57 $ & $0.85$ & $4.3\times 10^4$ & $19$ & $0.33$ & $0.83$ \\
$13$ & $115$ & $0.79$ & $1.2\times 10^5$ & $36$ & $0.31$ & $0.90$ \\
$14$ & $228$ & $0.72$ & $3.2\times 10^5$ & $60$ & $0.26$ & $0.94$ \\
$15$ & $431$ & $0.65$ & $8.7\times 10^5$ & $86$ & $0.20$ & $0.93$ \\
\hline
\end{tabular}
\caption{\textbf{Conf\/irmation of well-known results about RNA secondary structures.} For sequence length $L$, the table
lists the number of non-trivial structures $n_S$, the fraction of sequences $r_{triv}$ that fold into the trivial
structure, the average neutral network size $V_S$, the number of frequent structures $n_{freq}$, and the proportion
$r_{freq}$ of sequences in frequent structures to all sequences with a non-trivial structure.}
\label{tab:OldResults}
\end{table*}

\par In Table \ref{tab:Components} we show similar results, focusing on neutral components rather than networks. Just as
the number of NNs ($n_S$ in Table \ref{tab:OldResults}) the total number of NCs $n_C$ increases with sequence length.
More interesting is the result that the average number of components per network $n_C/n_S$ also increases with $L$. It
is worth noting that (for the range of $L$ we studied here) the mean number of components per network roughly doubles
when 2 more bases are added to the sequence. Crudely speaking, two more bases allow to form an extra base pair. This
rough argument then agrees nicely with our claim from the main text that the number of NCs can be expected to scale as
$2^n$ (where $n$ is the number of base pairs in the structure).

\par
\begin{figure}
\centering
\includegraphics[width=\figurewidth]{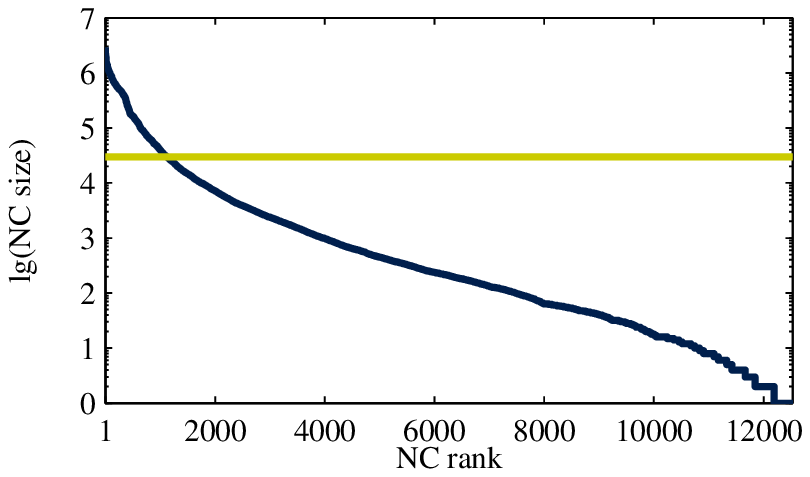}
\caption{\textbf{The distribution of NC sizes is even more skewed than for NNs.} Here the NCs are ranked by size,
again starting at rank $1$ for the largest NC. The yellow line indicates the average size. $1120$ NCs (less than
$10\%$) are larger than average, but together they cover $95\%$ of non-trivial genotype space (see Table
\ref{tab:Components}).}
\label{fig:nc_skewness}
\end{figure}

 While the distribution of NN sizes is heterogeneous (cf. Figure \ref{fig:nn_skewness}), the distribution of NC
sizes shows an even more pronounced skew (see Fig. \ref{fig:nc_skewness}). Again def\/ining an NC to be large if it is
greater than the average NC, the fraction of large NCs is smaller than the fraction of large NNs. Nonetheless, this
smaller proportion of components contains an even larger fraction of genotypes than the large NNs.

\begin{table*}[!htb]
\centering
\begin{tabular}{|r|r|r|r|r|r|r|}
\hline
\multicolumn{1}{|c}{$L$} & \multicolumn{1}{|c}{$n_C$} & \multicolumn{1}{|c}{$n_C/n_S$} & \multicolumn{1}{|c}{$\langle
V_C\rangle$} & \multicolumn{1}{|c}{$n_{lrg}$} & \multicolumn{1}{|c}{$n_{lrg}/n_C$} & \multicolumn{1}{|c|}{$r_{lrg}$} \\
\hline
$12$ & $641$ & $11$ & $3801$ & $133$ & $0.21$ & $0.87$ \\
$13$ & $1757$ & $15$ & $7998$ & $289$ & $0.16$ & $0.92$ \\
$14$ & $4603$ & $20$ & $16235$ & $580$ & $0.13$ & $0.94$ \\
$15$ & $12526$ & $29$ & $29972$ & $1120$ & $0.09$ & $0.95$ \\
\hline
\end{tabular}
\caption{\textbf{Overview of results for NCs.} For sequence length $L$, the table lists the number of NCs $n_C$, the
mean number of NCs per NN, the average NC size $\langle V_C\rangle$, the number of large NCs $n_{lrg}$, the
fraction of large NCs, and the fraction of sequence space occupied by large NCs, $r_{lrg}$.}
\label{tab:Components}
\end{table*}

\par
\begin{figure}[!ht]
\centering
\includegraphics[width=\figurewidth]{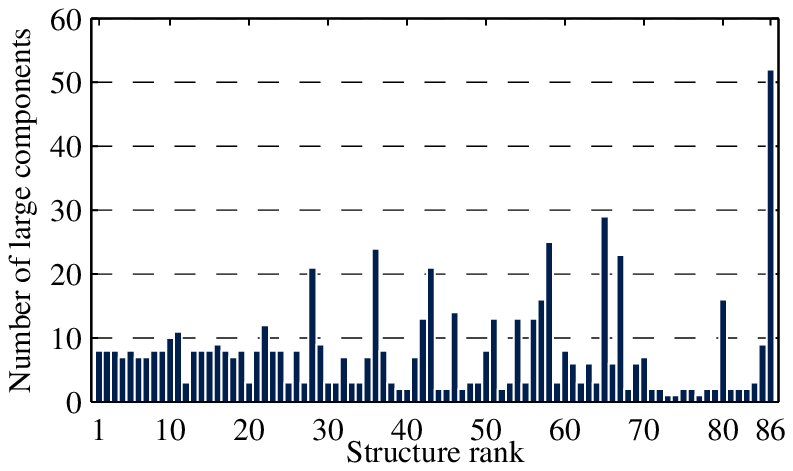}
\caption{\textbf{Most NNs contain several large NCs.} An NC is called large if its size is at least the average NC
size of its NN. The figure shows number of large NCs for the frequent structures of $L=15$. Only $3$ of them contain a
single large NC. When the rare structures are also counted, $56$ out of $431$ NNs contain only one large NC.}
\label{fig:large_ncs}
\end{figure}

\begin{figure}[!ht]
\centering
\includegraphics[width=\figurewidth]{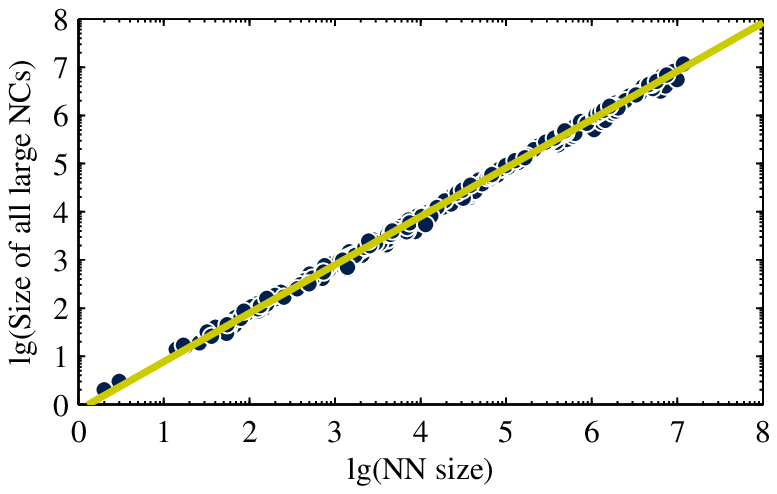}
\caption{\textbf{NNs are dominated by the large NCs.} The $x$-axis gives the size of a NN and the $y$-axis marks the
combined size of all large NCs of that NN. The yellow line is a straight-line least-squares f\/it.}
\label{fig:all_large_ncs}
\end{figure}

 Overall, these global considerations indicate a strong heterogeneity in genotype space. Does this heterogeneity
also exist within individual NNs? In Figure \ref{fig:large_ncs} we show the number of large NCs for all frequent NNs for
$L=15$; from now on, we call an NC large if it is greater than the average NC \emph{in its NN}. Almost all frequent NNs
contain several large NCs, only the ones ranked 73rd, 74th and 77th are dominated by a single large NC. Overall, for
$L=15$ there are 56 NNs with only a single large NC. 19 of these NNs are fully connected. It is clear that all NNs are
dominated by their large NCs (Figure \ref{fig:all_large_ncs}).

\par
\begin{figure}[!ht]
\centering
\includegraphics[width=\figurewidth]{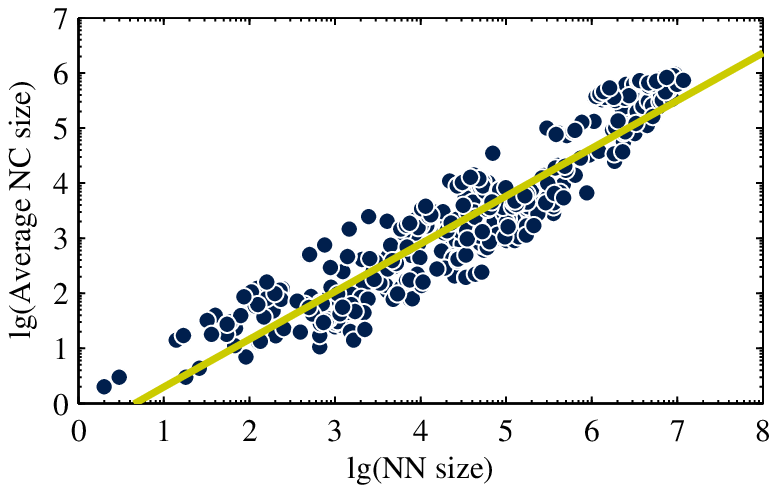}
\caption{\textbf{The NN size does not reliably predict average NC size.} As a consequence of the large variation in
the absolute number of NCs in an NN, it is possible that a smaller NN has larger NCs on average. The yellow line is a
straight line least squared f\/it to the data points.}
\label{fig:average_nc}
\end{figure}

\begin{figure}[!ht]
\centering
\includegraphics[width=\figurewidth]{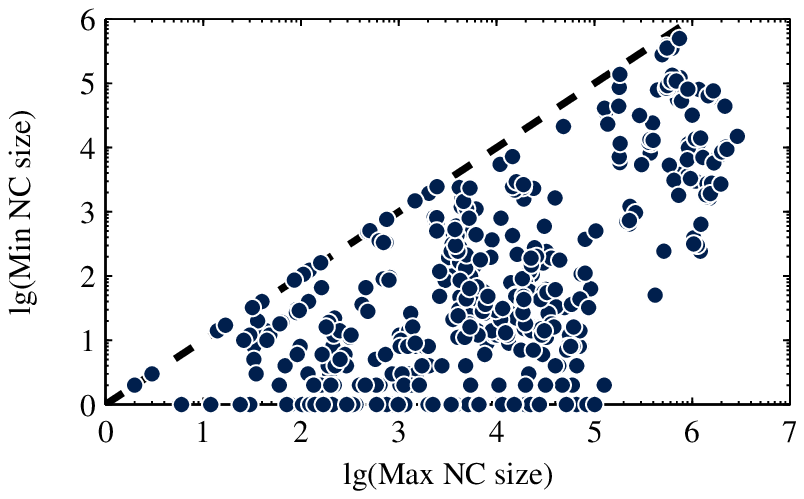}
\caption{\textbf{The scatter of NC sizes can be very large.} Each point in the figure corresponds to a NN and marks
the size of the largest and smallest NCs in that NN. The black dashed line indicates the equality of largest and
smallest NC; only fully connected NNs fall onto this line.  This result implies that the robustness of a phenotype can
vary widely depending on which neutral network it corresponds to.}
\label{fig:minmax}
\end{figure}

 Do larger NNs generally have larger NCs? Figure \ref{fig:average_nc} shows that this is not strictly the case. Of
course, the possible NC size is limited by the size of the entire NN. However, the number of NCs in an NN is not
strongly correlated with the size of the NN (see main text, Figure 1), but with the number of bonds in the corresponding
structure. Therefore, NNs of similar size can have quite different numbers of NCs. Thus the average NC size is not a reliable indicator of the size of
the corresponding NN. Additionally, the overall spread in NC sizes in each NN can be very large (Figure
\ref{fig:minmax}).

\par
\begin{figure*}[!htb]
\centering
\includegraphics[width=\widefigurewidth]{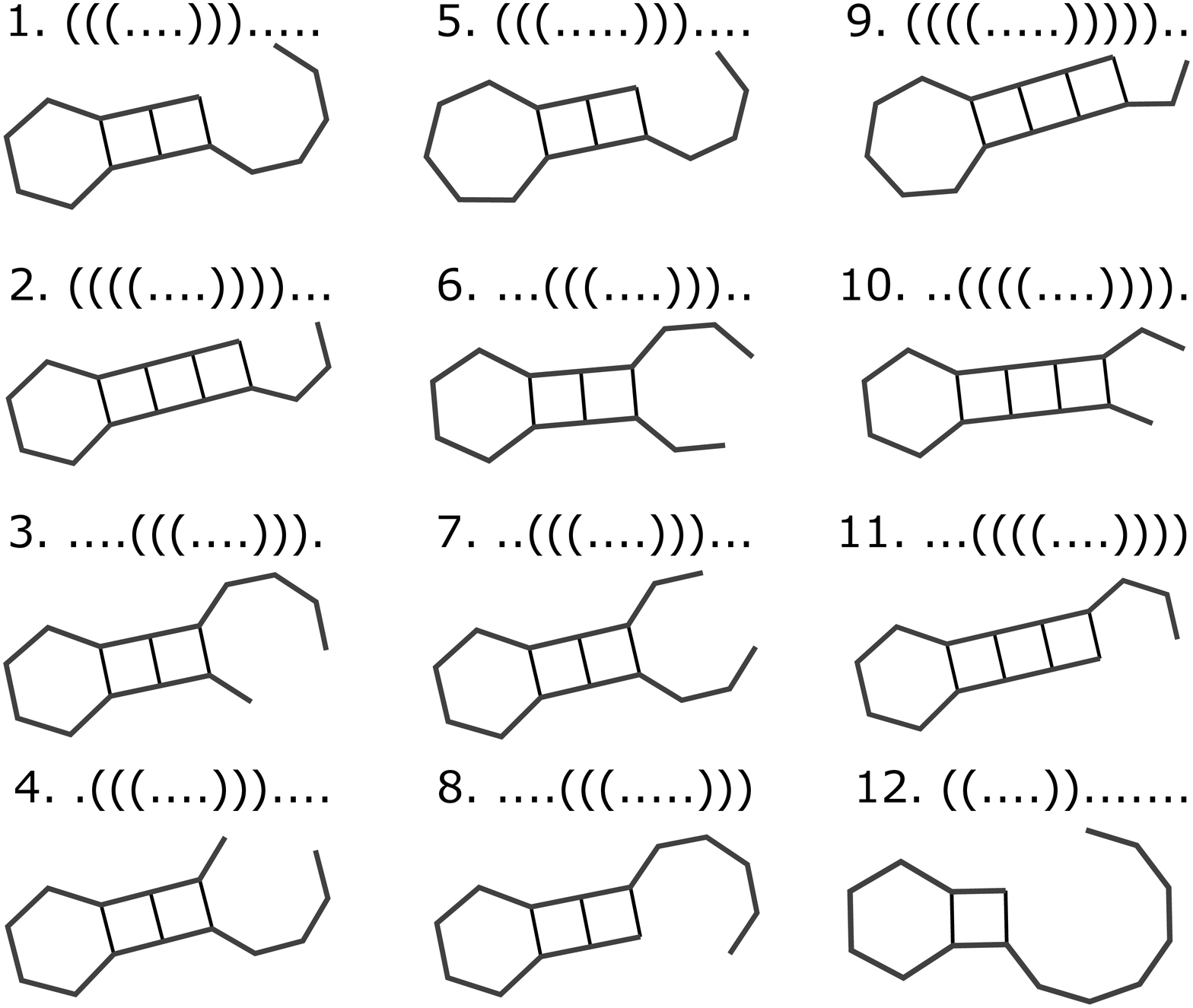}
\caption{\textbf{The most abundant secondary structures.} For each rank, the structure is given in the dot-bracket
notation and in a simple diagrammatic representation \cite{ViennaPackage}. The ribose backbone is drawn in grey and base
pairs are indicated by black lines. The structures follow counter-clockwise the direction from $5^\prime$ to
$3^\prime$.}
\label{fig:largest_structures}
\end{figure*}

 Due to the strong heterogeneity in the number of genotypes per  phenotype, the largest NNs are going to dominate genotype
space. In Figure \ref{fig:largest_structures}, we show the 12 most abundant structures; Table
\ref{tab:largest_structures} lists some of their properties. In particular, the number of (large) components is
identical (or close to) $2^n$, where $n$ is the number of base pairs in the structure. For some structures (e.g. the
second most abundant one), there are exactly the $2^n$ NCs we expect due to base pair exchanges (see the main
paper for details). 

\par This simple dependence of the number of components on the structure corresponding to the NN is striking. In fact,
it can be linked to the percolation theory arguments given by Reidys \cite{reidys1997percolation}. He considers random
GP maps and derives a threshold for the average number of neutral neighbour genotypes in order for an NN to percolate.
It is clear that the assumption of a random GP map cannot capture the neutral reciprocal sign epistasis that leads to NN
fragmentation in RNA. To accommodate for this effect, Reidys introduced another mutational move, namely base pair swaps:
In addition to point mutations of individual bases, paired bases are allowed to change in synchrony and thereby maintain
a bond.

\par When we stick with the more restrictive def\/inition of mutations as single nucleotide substitutions, it is still
possible to appeal to the percolation theory arguments. However, we need to restrict the set of genotypes to consider.
Specif\/ically, given a secondary structure we need to f\/ix for each bond whether the base pair is made in the order
purine-pyrimidine, or vice-versa. Thus we take only a subset of all possible genotypes into account. Within this subset,
we can expect the percolation argument to be applicable. So if a phenotype is suff\/iciently frequent (this notion is
made precise in \cite{reidys1997percolation}), the genotypes mapping into that phenotype will percolate \emph{in each
separate subspace}.

\par There are also biophysical reasons that may change our expectation of $2^n$ NCs. In particular, our results rest on
the assumption that a $GC$ base pair can be transformed into an $AU$ pair via a $GU$ intermediate. Empirically, at
our temperature of interest ($37^\circ$C) the Vienna package \cite{ViennaPackage} indicates that neutral intermediates
exist, at least in the frequent structures. Nonetheless, it may well be that these intermediates are not neutral in
reality. For example, Meer et al. \cite{kondrashov2010RNA} point out that $GU$ intermediates of tRNAs may have
a strong selective disadvantage. Our simple model in which neutrality is based only on identical secondary structure
cannot capture these effects. However, in principle it would be possible to take other factors such as free energy and
stability of the native fold into account.

\begin{table}[!hbt]
\centering
\begin{tabular}{|r|r|r|r|r|r|}
\hline
\multicolumn{1}{|c}{$R$} & \multicolumn{1}{|c}{$V$} & \multicolumn{1}{|c}{$2^n$} &
\multicolumn{1}{|c}{$N_c$} & \multicolumn{1}{|c}{$N_{lrg}$} & \multicolumn{1}{|c|}{$r$} \\
\hline
1 & 11795379 & 8 & 16 & 8 & 0.53 \\
2 & 9978003  & 16 & 16 & 8 & 0.67 \\
3 & 9454721  & 8 & 10 & 8 & 0.68 \\
4 & 8988572  & 8 & 10 & 7 & 0.58 \\
5 & 8698911  & 8 & 26 & 8 & 0.49 \\
6 & 8303219 & 8 & 10 & 7 & 0.57 \\
7 & 8050101  & 8 & 10 & 7 & 0.56 \\
8 & 8001910 & 8 & 15 & 8 & 0.46 \\
9 & 7914436  & 16 & 16 & 8 & 0.56 \\
10 & 7675391  & 16 & 16 & 10 & 0.69 \\
11 & 7647918  & 16 & 20 & 11 & 0.31 \\
12 & 7525506  & 4 & 9 & 3 & 0.15 \\
\hline
\end{tabular}
\caption{\textbf{Overview of the most frequent structure for $L=15$.} Structures are ranked by their size $V$ (the
number of genotypes mapping into them) starting at rank $R=1$ for the most frequent structure. $n$ is the number of
bonds in the structure (cf. Figure \ref{fig:largest_structures}). $N_c$ is the total number of NCs in the corresponding
NN, and $N_{lrg}$ is the number of large NCs. $r$ is the size ratio of the $2^n$th NC to the largest NC.}
\label{tab:largest_structures}
\end{table}

\newpage\par\qquad
\newpage\par\qquad

\subsection{Crossover does not provide compensatory mutations}
\begin{figure*}[!ht]
\centering
\includegraphics[width=\widefigurewidth]{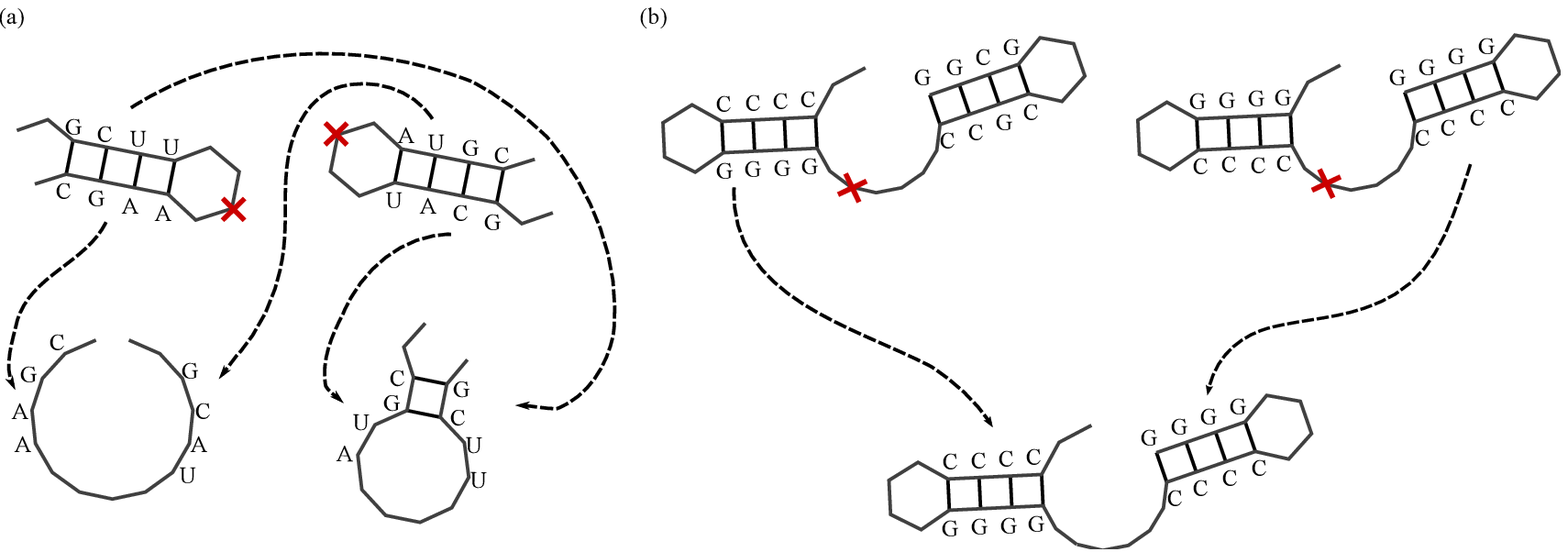}
\caption{\textbf{Illustration of cross-over.} (a) Mismatches in the same stack break the structure. (b) Mismatches in
different stacks can lead to the discovery of a new component. In both cases, the red cross indicates the cross-over
point.}
\label{fig:crossover}
\end{figure*}

\par In our construction of the genotype space hypercube, we have only taken point mutations into account when we
connected neighbouring genotypes. In nature, other mutational moves are also observed: Deletions or insertions
correspond to omitting or adding bases during reproduction of the genotype. These mutations change the sequence length
which makes them hard to discuss in our framework.

\par There is one other kind of mutation that we can address, namely crossover. This mutational move is particularly
relevant to sexually reproducing organisms in which the offspring receives part of its genotype from each
parent\footnote{Asexual organisms such as bacteria can also mutate under crossover by horizontal gene transfer}. Can
crossover connect separate NCs?

\par The simplest case to study is when both parental genotypes belong to the same NC. Let us focus on a single base
pair for simplicity -- if any single pair is not maintained in the offspring phenotype, that phenotype is necessarily
different from the parental one. For def\/initeness, we assume that the base pair of interest is made by a
purine-pyrimidine pair in both parents, say GC in one and AU in the other\footnote{Having a GC and a GU pair in the
same NC is likely to occur only in frequent structures. For rare phenotypes, only GC may exist; this is not important
to the argument}. Evidently, crossover between the parents will again result in a purine-pyrimidine pair. Depending on
the parental base pairs, this could be GC, GU, AU or AC. The f\/irst three pairs are compatible and may thus lead to the
parental phenotype. However, they will again be part of the same NC -- a transition into a pyrimidine-purine pair (which
would necessarily be part of a different NC) is not possible. Finally, an AC pair is incompatible and will lead to a
different phenotype.

\par The case of parental genotypes from different NCs is slightly more involved. Let us again consider a
purine-pyrimidine pair in the f\/irst parent, but now a pyrimidine-purine pair in the other parent. We now need to
distinguish two cases depending on the point of crossover. First, consider the case that the crossover point is within
the stem-loop region enclosed by the base pair of interest (illustrated in Figure \ref{fig:crossover}a). This means that
the resulting genotypes will have either a purine-purine or a pyrimidine-pyrimidine pair, neither of which can form a
bond. Thus the offspring phenotype is necessarily different from the parent. Second, the point of crossover may be
outside the stem-loop region of interest. In that case, the base pair is left intact. If the parental phenotype contains
two separate stem-loop regions, this scenario of crossover may lead to a new NC: Individually incompatible stems are
collectively `shuffled'. Yet this way of exploring different NCs is limited to reusing the already existing stems; new
variants of individual stems cannot be achieved in this manner.

\par In summary, crossover alone cannot lead to new NCs. In order to create genotypes on an NC that is different from
the parental NCs, it is necessary to cross genotypes from different NCs in special positions. However to arrive at a new
NC in the f\/irst place still requires two mutations even if crossover is taken into account.

\newpage\par\qquad

\section{Neutral components shape evolutionary trajectories}

\par
\begin{figure}[!ht]
\centering
\includegraphics[width=\figurewidth]{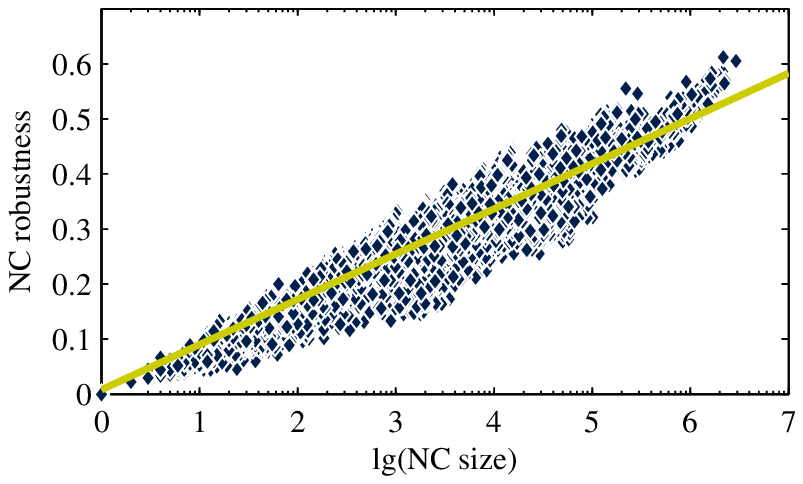}
\caption{\textbf{NC robustness increases with size.} The robustness of an NC is the average normalized connectivity of the
genotypes in the NC; the standard deviation is always much smaller than the average, so that the average is a
meaningful quantity (data not shown). The blue points are the actual data, the yellow line is a straight line least squared f\/it.}
\label{fig:robustness}
\end{figure}

\begin{figure}[!ht]
\centering
\includegraphics[width=\figurewidth]{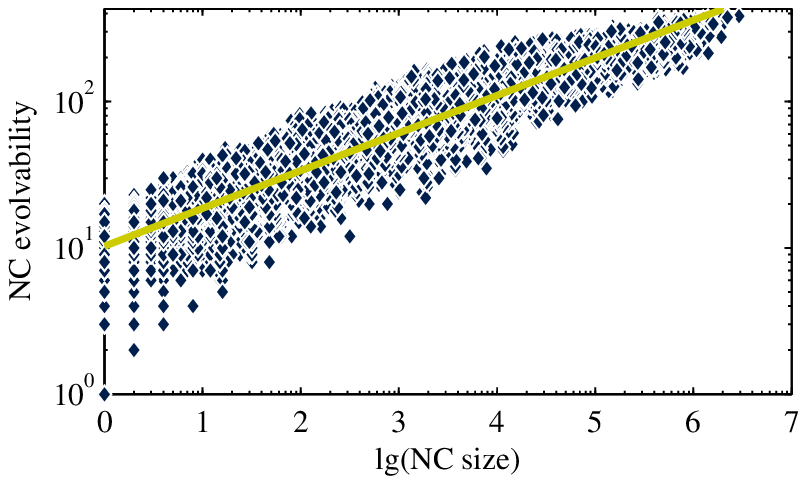}
\caption{\textbf{The number of accessible phenotypes increases with size.} We measure evolvability of an NC by counting
the number of distinct phenotypes that can be reached by a point mutation off some genotype in the NC. The blue points are the data,
the yellow line is a linear f\/it to the log-log data. The black dashed line indicates the total number of structures
($431$ for $L=15$).}
\label{fig:evolvability}
\end{figure}

 In the main paper, Fig. 3 we show that the robustness to genetic change and the number of phenotypes in the
one-mutant neighbourhood of an NC and its size are correlated. Here, we provide additional views of this data set,
focusing on the correlation of robustness and NC size (Figure \ref{fig:robustness}) and on the correlation of
evolvability and NC size (Figure \ref{fig:robustness}).

\subsection{Common and joint evolvabilities for large NNs} 

\begin{figure}[!ht]
\centering
\includegraphics[width=\figurewidth]{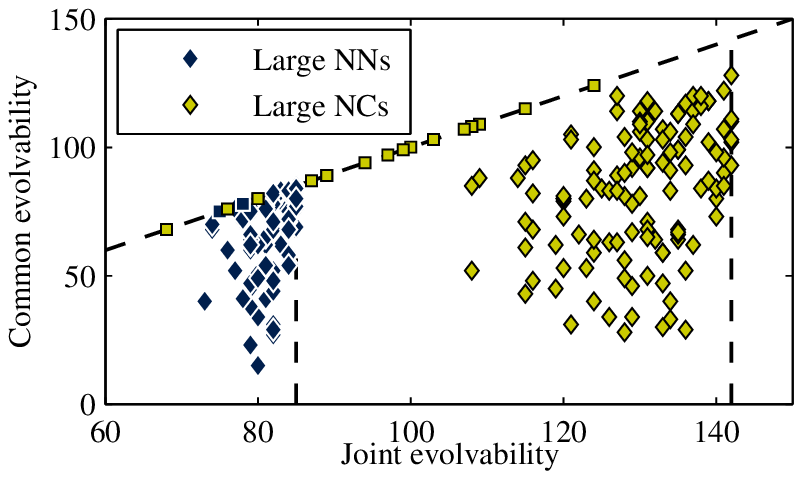}
\caption{\textbf{Even large NCs are not homogeneously connected.} Joint and common evolvability are calculated
taking only large NCs into account. Blue markers correspond to including on the large NCs of the large NNs; yellow
markers are for the case of all large NCs (see text for further explanation). Square markers indicate NNs with only a
single large NC; the vertical dashed lines show the number of NNs for each calculation, which correspond to the maximum
evolvability. The other dashed line shows the equality of the joint and common evolvabilities.}
\label{fig:joint_common_large}
\end{figure}

\par Given the large skew in the size of the NNs, it is worth considering another question regarding joint and common
evolvability: Can the large NNs be reached from each other?  By only considering the large NCs of the large NNs, we
account for $86$ NNs and $78\%$ of (non-trivial) genotype space. Alternatively, we can consider all NCs which are
larger than the average NC size (calculated from all NCs, not just a particular structure). There are $143$ NNs with at
least one such NC, and all $1120$ large NCs together cover $95\%$ of all non-trivially folding genotypes. In both
cases, we observe in Fig. \ref{fig:joint_common_large} that the discrepancy between joint and common evolvability
remains signif\/icant.

\subsection{Relative evolvability for individual NCs}

\par
\begin{figure*}[!ht]
\centering
\includegraphics[width=\widefigurewidth]{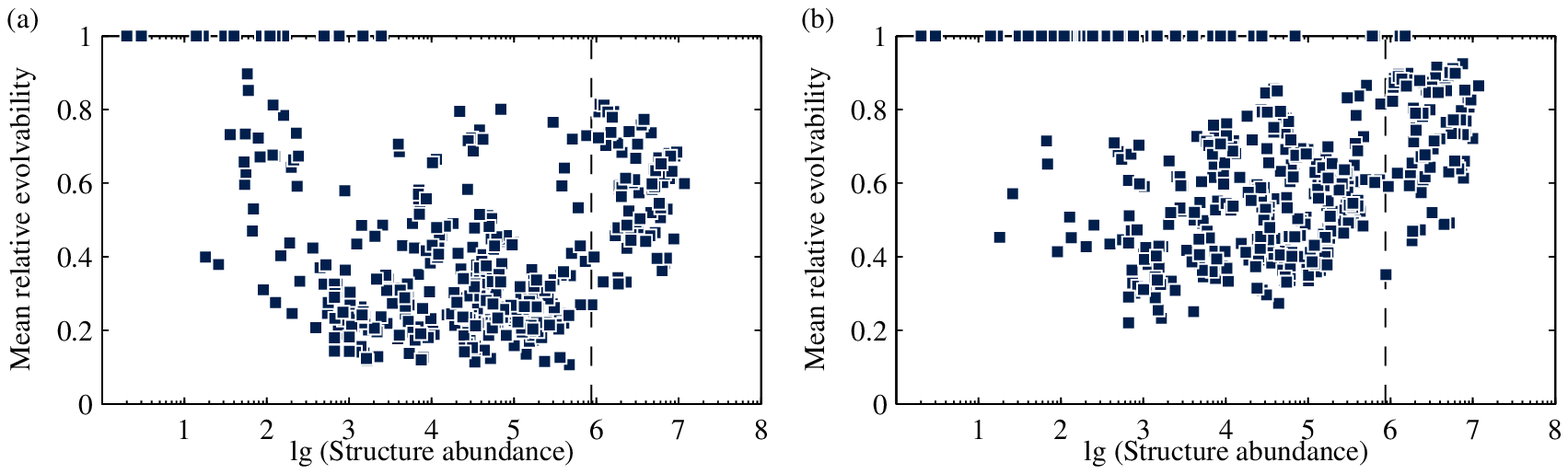}
\caption{\textbf{Relative evolvability of individual NCs.} The relative evolvabiltiy of an NC is def\/ined as the
evolvability of that NC divided by the joint evolvablity of the NN to which the NC belongs. \textbf{(a)} Here we
calculate the relative evolvability for each NN by averaging over \emph{all} NCs. \textbf{(b)} Here the average is computed
for the \emph{large} NCs of each NN only, increasing the mean relative evolvability. In both panels, the vertical
black dashed line marks the average NN size.}
\label{fig:rel_evo}
\end{figure*}

 The common evolvability (as def\/ined in the main text, Eqn. (3)) gives the number of phenotypes that can be reached
from any NC in a given NN. This is often much less than the joint evolvability (main text, Eqn. (2)) indicating that there
are many phenotypes that can only be reached from some, but not all NCs with a given phenotype. But this does not tell
us how many phenotypes can be reached on average from a NC in a given network. To measure this quantity, we def\/ine the
\emph{mean relative evolvability} as the ratio of NC evolvability to NN (joint) evolvability, averaged over all NCs in
the network. Alternatively, we can restrict the average to the large NCs only. In Figure \ref{fig:rel_evo}, we show that
this fractional evolvability increases slightly with phenotype abundance, but remains clearly below unity. This means
that to sample all phenotypes that are part of the joint evolvability, it is necessary for a population to jump between
NCs.

\par If we take the average over all NNs (cf. Figure \ref{fig:rel_evo}a), we f\/ind that the mean relative evolvability is
around $42\%$; for the frequent NNs, this average is $59\%$. If we take only large NCs into account (Fig.
\ref{fig:rel_evo}b), the average is $63\%$ for all NNs and $76\%$ for the frequent NNs.

\subsection{Choosing large NCs according to entropy stresses the importance of fragmentation}
\label{sec:entropy}
\par Our def\/inition of what constitutes a large NC (namely it must be larger than the average NC its NN) is in analogy
with we call a frequent phenotype \cite{gruener1996RNAexhaustive}. An alternative def\/inition can be made in terms of the
entropy of the distribution of NC sizes. If we denote the relative size of NC $i$ by $f_i$ we have $\sum_i f_i = 1$
where the sum is over the NCs of a given NN. The entropy of the distribution is then 
$$S=-\sum_i f_i \log f_i$$
If the NN in question were fragmented into $N$ NCs of equal size, we would obtain $f_i = 1/N$ and $S = \log N$. So
$\exp(S)$ gives us an approximate number of NCs whose relative size is signif\/icant in the NN. So in order to determine
the large NCs in a given NN, we calculate $\exp(S)$, round to the nearest integer $N$ and choose the $N$ largest NCs of
the NN.

\par In general, the entropy requirement is less restrictive than choosing the average size as a threshold: For only $6$
NNs (with ranks between 80 and 159) the number of large NC is reduced under the entropy criterion (by 1 NC each). For 58
NNs, the number of NCs is the same for both criteria. Thus there remain 367 NNs which have more large NCs by entropy
than by average size, and the increase can be up to 5-fold (on average, it is 1.6-fold). In absolute terms, the entropy
criterion produces 4 additional large NCs on average.

\par
\begin{figure}[!ht]
\centering
\includegraphics[width=\figurewidth]{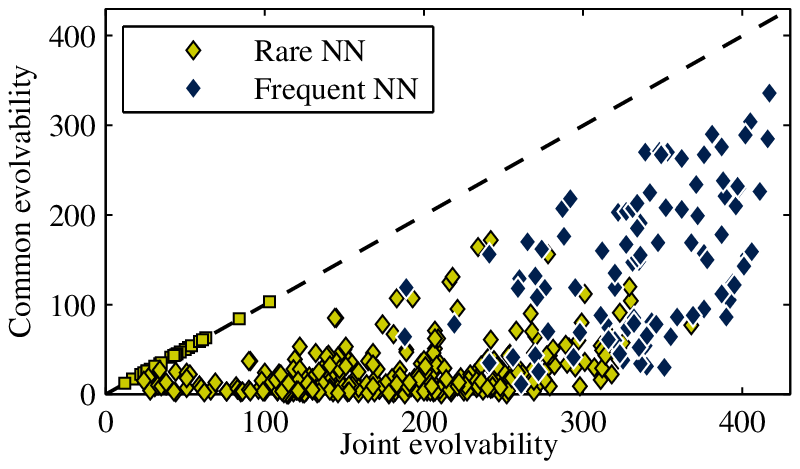}
\caption{\textbf{The distinction between joint and common evolvability is robust to the choice of large NCs.} In
analogy to Fig. \ref{fig:joint_common_evolvability}b in the main paper, the joint and common evolvability are
calculated for the large NCs of each NN. However, in this figure the entropy criterion was used to determine the large
NCs.}
\label{fig:evo_by_entropy}
\end{figure}

 Regarding the discrepancy between joint and common evolvability, it is clear that a larger number of NCs cannot
have a smaller joint evolvability or a bigger common evolvability than a smaller set. So if we consider the joint and
common evolvability of the large NCs in an NN, the entropy measure will - for almost all NCs - increase the gap between
the two values. This is illustrated in Fig. \ref{fig:evo_by_entropy}. The ratio of common to joint evolvability, when
calculated for the NCs that are large according to the entropy criterion, is $F_{S}=0.25$ when averaged over all NNs.
$F_S$ correlates with NN size: $r=0.28, p<10^{-8}$, so the discrepancy between joint and common evolvability is less
pronounced for large NNs. All these results are quantitatively similar and qualitatively consistent with the average
size criterion for large NCs that we have adopted in the main paper.

\par Given the two different criteria for what constitutes a large NC, is one more appropriate than the other?
One advantage of the entropy requirement is that it does not introduce a somewhat arbitrary, hard cutoff. Choosing the
average NC size as a threshold means that it is practically impossible to f\/ind that all NCs are large - this would arise only
if all NC had \emph{exactly} the same size. It thus appears that the entropy requirement should be favoured. Clearly,
this measure is more lenient in that on average it classif\/ies more NCs as being large. As our main interest in this
paper is to discern whether the fragmentation of NNs is important for evolutionary dynamics, we have chosen the more
restrictive, average-based approach in the main paper. This criterion on average produces less large NCs; thus the
effect of fragmentation is less pronounced. Thus the more natural, entropy-based approach suggests that NN
fragmentation could be even more severe than outlined in our paper.

\newpage\par\qquad
\section{Sampling at $L=20$ conf\/irms our results}
\label{sec:l20}
\par As genotype space grows exponentially with sequence length, exhaustive enumeration becomes infeasible for longer
(and thus biologically more interesting) sequences. In particular, to study the evolvability of NCs we need to
explore them completely. For RNA NNs, we can exploit the fact that their fragmentation has a simple cause, namely base
pair complementarity. Using this insight, it is relatively straightforward to obtain sequences with the same structure
but from different NCs: The Vienna package \cite{ViennaPackage} includes a function to inverse-fold structures with
constraints. Thus we can f\/ix the base pairs and then try to obtain sequences with the desired structure. Using these
sequences the NCs of a given NN can be mapped out. The evolvability can be calculated exactly for each NC by simply
keeping track of the phenotypes found by non-neutral mutations.

\par There is no guarantee that this approach will f\/ind all NCs; clearly, we are more likely to discover sequences on
the large NCs. For computational convenience, we have run the inversion algorithm on $100$ sequences for each of the
$6^n$ conf\/igurations of paired bases (again, $n$ is the number of base pairs), choosing the unpaired base uniformly at
random for each attempt. We should thus be able to f\/ind NCs of sizes ranging over at least $2$ orders of magnitude; in fact,
this range has turned out much larger, getting up to $6$ orders of magnitude. Details of the structures we sampled are
given in Tab. \ref{tab:sampled}.

\par In order to demonstrate that we found most genotypes of an NN, we use the sampling algorithm by J\"org et al.
\cite{joerg2008NNsize} to estimate NN sizes. By comparing the estimated NN size to the number of genotypes found by our
sampling approach, we f\/ind an indication whether the sampling approach has found all major NCs. 

\par It is important to be clear what it means that smaller NCs may not be found. From the def\/initions (Eqns. ($1$) and
($2$) in the main text) it is evident that any subset of NCs gives us a \emph{lower bound} on the joint evolvability and
an \emph{upper bound} on the common evolvability. Therefore, if the sampling is incomplete, better results can only show
that the discrepancy between $E^{(j)}$ and $E^{(c)}$ is greater than our results indicate.

\par
\begin{figure*}[!ht]
\centering
\includegraphics[width=\widefigurewidth]{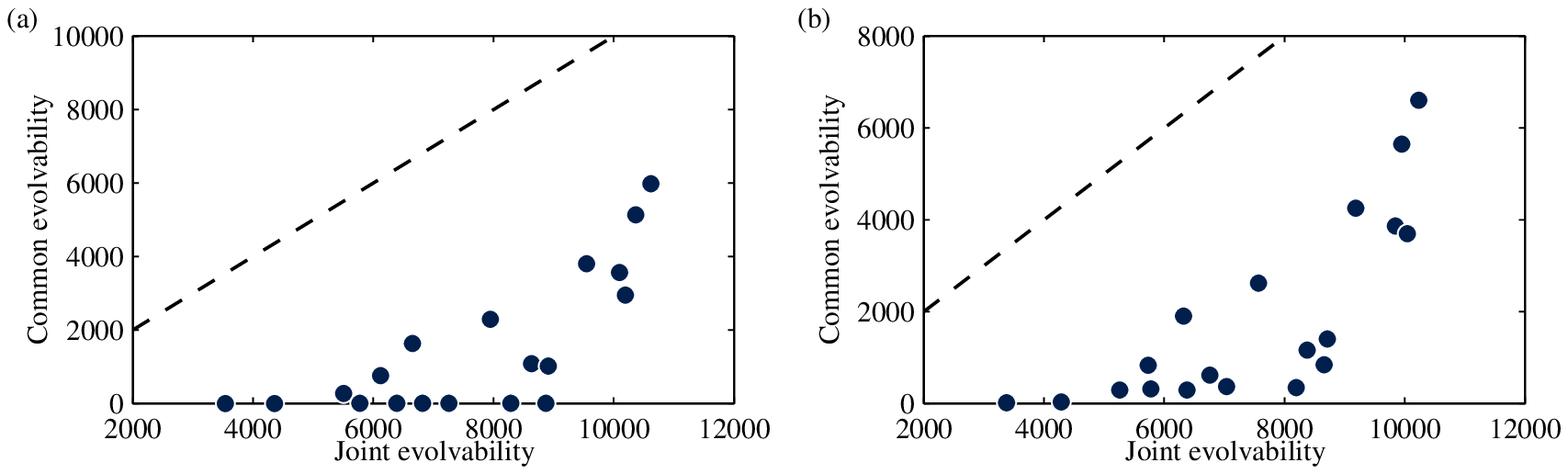}
\caption{\textbf{NC heterogeneity increases with sequence length}. Shown are the joint and common evolvability (as
def\/ined in the main text) of the structures sampled, according to Tab. \ref{tab:sampled}. In (a) all NCs have been
included while for (b) only NCs of more than average size have been used. }
\label{fig:sampling_evo}
\end{figure*}

 In general, we f\/ind that the discrepancy between joint and common evolvability is large (Fig.
\ref{fig:sampling_evo}): on average, $F \equiv E^{(c)}/E^{(j)} = 0.16$ and $F_{large} = 0.21$. Therefore, the
contingency due to neutral space fragmentation will be important at biologically realistic sequence lengths.

\begin{table*}[!hbt]
\centering
\begin{tabular}{|c|c|r|r|r|r|r|}
\hline
Structure & Estimated size & \multicolumn{1}{|c}{Sampled size} & \multicolumn{1}{|c}{NCs} & 
\multicolumn{1}{|c}{Large NCs} & \multicolumn{1}{|c}{$2^n$} & \multicolumn{1}{|c|}{$r$} \\
\hline
\texttt{((((....))))........} & $7.4\times 10^9 \pm 2.3\times 10^8$ & $7.4\times 10^9$ & $16$ & $7$  & 16 & 0.65 \\
\texttt{.((((....)))).......} & $5.4\times 10^9 \pm 2.0\times 10^8$ & $5.5\times 10^9$ & $16$ & $8$  & 16 & 0.62 \\
\texttt{..((((....))))......} & $4.8\times 10^9 \pm 2.2\times 10^8$ & $4.8\times 10^9$ & $16$ & $9$  & 16 & 0.58 \\
\texttt{(((((....)))))......} & $4.4\times 10^9 \pm 9.1\times 10^7$ & $4.4\times 10^9$ & $32$ & $19$  & 32 & 0.70 \\ 
\texttt{...((((....)))).....} & $4.2\times 10^9 \pm 2.4\times 10^8$ & $4.1\times 10^9$ & $16$ & $9$  & 16 & 0.54 \\
\texttt{....((((....))))....} & $3.9\times 10^9 \pm 1.9\times 10^8$ & $4.0\times 10^9$ & $16$ & $9$  & 16 & 0.53 \\
\texttt{......(((((....)))))} & $3.6\times 10^9 \pm 7.9\times 10^7$ & $3.6\times 10^9$ & $40$ & $24$  & 32 & 0.28 \\ 
\texttt{...(((((....)))))...} & $2.4\times 10^9 \pm 6.6\times 10^7$ & $2.5\times 10^9$ & $32$ & $18$  & 32 & 0.56 \\ 
\texttt{((((((....))))))....} & $2.0\times 10^9 \pm 3.3\times 10^7$ & $2.0\times 10^9$ & $64$ & $40$ & 64 & 0.66 \\ 
\texttt{....((((((....))))))} & $1.7\times 10^9 \pm 2.5\times 10^7$ & $1.7\times 10^9$ & $80$ & $47$  & 64 & 0.30 \\ 
\texttt{..((((((....))))))..} & $1.3\times 10^9 \pm 3.0\times 10^7$ & $1.3\times 10^9$ & $64$ & $38$  & 64 & 0.65 \\
\texttt{..((((((....)).)).))} & $2.8\times 10^8 \pm 1.5\times 10^7$ & $2.7\times 10^8$ & $117$ & $62$  & 64 & 0.24 \\ 
\texttt{((((..((....))..))))} & $2.7\times 10^8 \pm 1.0\times 10^7$ & $2.7\times 10^8$ & $115$ & $55$  & 64 & 0.24 \\
\texttt{((....))..(((....)))} & $4.9\times 10^7 \pm 7.4\times 10^6$ & $4.4\times 10^7$ & $138$ & $29$  & 32 & 0.08 \\ 
\texttt{((.((.((....))))..))} & $3.4\times 10^7 \pm 2.6\times 10^6$ & $3.9\times 10^7$ & $166$ & $65$  & 64 & 0.22 \\ 
\texttt{(((....)))..((....))} & $2.6\times 10^7 \pm 6.7\times 10^6$ & $2.2\times 10^7$ & $150$ & $36$  & 32 & 0.13 \\ 
\texttt{((....))....((....))} & $1.1\times 10^7 \pm 2.6\times 10^6$ & $1.1\times 10^7$ & $84$ & $18$ & 16 & 0.14 \\ 
\texttt{((.((.((....)).)).))} & $5.0\times 10^6 \pm 1.1\times 10^6$ & $4.7\times 10^6$ & $546$ & $105$  & 64 & 0.10 \\ 
\texttt{((((..((....))))..))} & $7.3\times 10^5 \pm 3.5\times 10^5$ & $6.4\times 10^5$ & $750$ & $100$ & 64 & 0.04 \\
\hline

\end{tabular}
\caption{\textbf{Details of the sampled structures.} For each structure, its representation in the dot-bracket notation
is given, together with the estimated and sampled sizes and the number of components. Large components are those that
are greater than the average of the NCs that have been found - note that when small NCs are not found, this produces an
over-estimate of the average. $n$ is the number of bonds in the structure, and $r$ is the ratio of the $2^n$th NC size
to the largest NC.}
\label{tab:sampled}
\end{table*}

\newpage\par\qquad
\newpage\par\qquad

\section{NCs beyond RNA}
\par
\begin{figure}[!ht]
\centering
\includegraphics[width=\figurewidth]{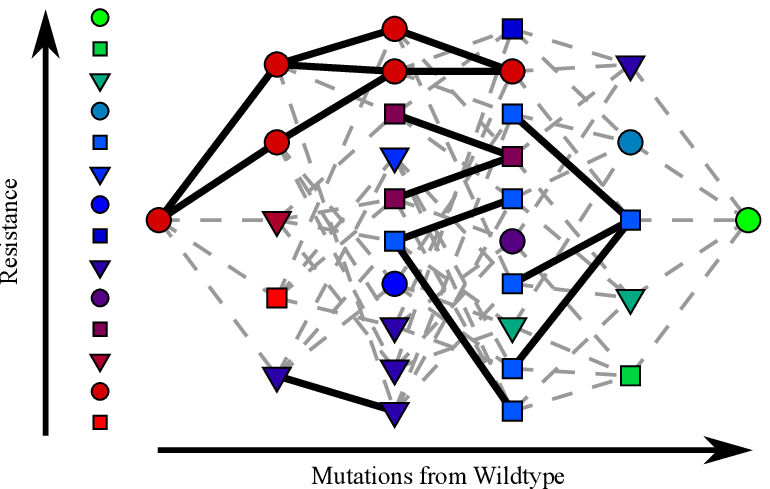}
\caption{\textbf{Experimental evidence for NCs}. The figures shows the genotype landscape investigated by Weinreich
et al. \cite{weinreich2006paths}. Neutral mutations are indicated by solid lines, non-neutral mutations are dashed. The
horizontal position of each marker indicates its mutational distance from the wild-type, and its shape and colour
indicate its f\/itness.}
\label{fig:weinreich}
\end{figure}

 Establishing the fragmentation of a NN is challenging. If we cannot exploit simple biophysical principles (as in
the case of RNA), or if we would like to study properties of NCs, we need to determine phenotypes for large portions of
genotype space. Due to the vast numbers of genotypes even in small systems, this is a demanding task. Experimental
progress in this area is underway; a pioneering study by Weinreich et al. \cite{weinreich2006paths} has determined the
antibiotic resistance due a particular $\beta$-lactamase. The authors considered 5 mutations that jointly increase the
resistance by about 100,000 fold and measured the resistance of all intermediate mutants between the wildtype and the
most resistant type. In Figure \ref{fig:weinreich}, we show the resistance landscape that arises in this system. Within
the resolution of the experiment, there are several NNs of genotypes that convey the same resistance. Some of these NNs
contain multiple NCs; however, it cannot be ruled out that mutations outside the scope of the experiment connect these NCs.

\par Another system in which the mapping from genotype to phenotype is well characterized is the genetic code, linking
triplets of nucleotides in DNA (or mRNA) to amino acids in proteins\footnote{In many other contexts, the sequence of
residues in a protein is considered as its genotype. Fundamentally however, mutations change DNA, while biological
function depends on the amino acid sequence.}. Here, a neutral network consists of all the codons that are translated
into the same amino acid (we treat the STOP signal as another amino acid). In the universal genetic there are 21 NNs with sizes
between 6 (arginine, leucine, serine) and 1 (methionine, tryptophan). Serine is the only amino acid whose NN is
fragmented. This NN contains NCs which have size 2 and 4, respectively. So in total, there are 22 NCs in 21 NNs.

\par
\begin{figure}[!htb]
\centering
\includegraphics[width=\figurewidth]{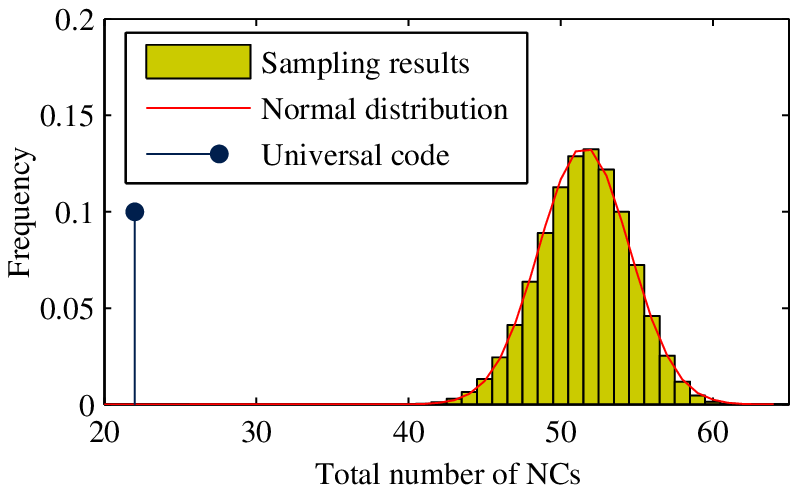}
\caption{\textbf{Random genetic codes show a high degree of NN fragmentation}. $4\times 10^7$ genetic codes were
generated by assigning each codon a random amino acid, keeping the degeneracy of the universal code, and the number
of NCs in each code were evaluated. The data has mean $\mu = 51.6$ and standard deviation $\sigma = 3.0$. The red line
shows a normal distribution with these parameters. The position of the universal genetic is indicated in blue. The
smallest number of components found in the sampled data is 33.}
\label{fig:genetic_code}
\end{figure}

 It is well known that the universal genetic code is signif\/icantly different from an arbitrary assignments of codons
to amino acids \cite{wagner2005book}. Here, we are interested to discern if the NN connectivity observed in the code is
another property that sets the universal code apart from a random alternative. To this end, we generated $4\times 10^7$
codes by assigning each codon a randomly chosen amino acid, such that the degeneracy of the universal code is
maintained. For each realization, we counted the overall number of NCs. A histogram of the data is shown in
Figure \ref{fig:genetic_code}. On average, a random code contains $51 \pm 3$ NCs, much more than the 22 NCs of the
universal code. The lowest number of NCs in our sample was 33 and this was realized only once. The maximum possible
number of NCs, 64, was found in 23 random realizations of the code.

\par It is clear that the degree of NN connectivity of the universal code is far from random. However, the universal
code does not minimize the total number of NCs completely -- for example, exchanging the 2 tyrosine codons with the
smaller NC of serine would yield a maximally connected code. One possible explanation of the high connectivity of the
universal code is that it confers robustness of the protein amino acid sequence to point mutations in the DNA and to
translation errors \cite{wagner2005book}. This f\/inding illustrates an important message of the main paper: The
robustness of a phenotype (the amino acid) cannot be ascribed to the properties of phenotype itself (such as the
degeneracy of the amino acid), but is sensitive to the local connectivity of the NC. 

\end{document}